\documentclass[aps,prl,twocolumn,showpacs,superscriptaddress,groupedaddress]{revtex4}  

\usepackage{amssymb}
\usepackage{amsmath}
\usepackage{epsfig}
\usepackage{epstopdf}
\usepackage{bm}
\usepackage{graphicx,epsfig}
\usepackage{mathrsfs}
\usepackage{dcolumn}
\usepackage{color}
\usepackage{natbib}
\usepackage{CJK}
\usepackage{miniplot}
\usepackage{subfigure}
\def\be{\begin{equation}}
\def\ee{\end{equation}}
\def\bea{\begin{eqnarray}}
\def\eea{\end{eqnarray}}
\def\la{\langle}
\def\ra{\rangle}

\allowdisplaybreaks[2]

\begin{document}

\title{Superradiance Lattice}
\author{Da-Wei Wang}
\email{cuhkwdw@gmail.com}
\affiliation{Texas A$\&$M University, College Station, TX 77843, USA}
\affiliation{Department of Physics and Centre for Quantum Coherence, The Chinese University of Hong Kong, Hong Kong, China}
\author{Ren-Bao Liu}
\email{rbliu@phy.cuhk.edu.hk}
\affiliation{Department of Physics and Centre for Quantum Coherence, The Chinese University of Hong Kong, Hong Kong, China}
\author{Shi-Yao Zhu}
\affiliation{Department of Physics and Centre for Quantum Coherence, The Chinese University of Hong Kong, Hong Kong, China}
\affiliation{Beijing Computational Science Research Centre, Beijing 100084, China}
\author{Marlan O. Scully}
\affiliation{Texas A$\&$M University, College Station, TX 77843, USA}
\affiliation{Princeton University, Princeton, New Jersey 08544, USA}
\affiliation{Baylor University, Waco, TX 76706, USA}

\date{\today }

\begin{abstract}
We show that the timed Dicke states of a collection of three-level atoms can form a tight-binding lattice in momentum space. This lattice, coined the superradiance lattice (SL), can be constructed based on electromagnetically induced transparency (EIT). For a one-dimensional SL, we need the coupling field of the EIT system to be a standing wave. The detuning between the two components of the standing wave introduces an effective uniform force in momentum space. The quantum lattice dynamics, such as Bloch oscillations, Wannier-Stark ladders, Bloch band collapsing and dynamic localization can be observed in the SL. The two-dimensional SL provides a flexible platform for Dirac physics in graphene. The SL can be extended to three and higher dimensions where no analogous real space lattices exist with new physics waiting to be explored.
\end{abstract}

\pacs{42.50.Nn, 61.50.Ah}

\maketitle

\emph{Introduction}.---Since the early days of quantum mechanics, the periodic lattice has been a platform for versatile quantum phenomena of electrons, such as Bloch oscillations \cite{Bloch1929, Zener1934}, Wannier-Stark ladders \cite{Wannier1960} and dynamic localization \cite{Ignatov1976, Dunlap1986}. Bloch oscillations and Wannier-Stark ladders have been observed in superlattices \cite{Leo1992, Lyssenko1997} and optical lattices \cite{BenDahan1996, Wilkinson1996}. The evidence of dynamic localization and Bloch band collapsing \cite{Holthaus1992} under periodic forces were also observed in optical lattices \cite{Graham1992, Moore1994, Madison1998} and photonic structures \cite{Szameit2010}. The related Floquet lattice phenomena include quantum phase transitions \cite{Eckardt2005, Lignier2007, Zenesini2009, Struck2011}, Majorana fermions \cite{Jiang2011, Liu2013}, topological insulators \cite{Kitagawa2010, Lindner2011, Hauke2012}, artificial gauge potentials \cite{Struck2012, Aidelsburger2013, Miyake2013, Goldman2014} and edge states \cite{Gomez-Leon2013, Rudner2013}. Apart from these nonrelativistic physics, the invention of graphene \cite{Novoselov2005} brought a new stage of relativistic Dirac physics \cite{CastroNeto2009} in two-dimensional lattices. Nevertheless, the observation of these phenomena remains challenging. Novel types of lattices \cite{Shapere2012, Wilczek2012, Li2012a, Guo2013} provide new testing grounds for the rich physics mentioned above.

In this Letter, we introduce the concept of the superradiance lattice (SL), a lattice in momentum space \cite{Cooper2013}. The conventional lattice has discrete translational symmetry in position space. The tight-binding model which allows electron hopping between nearest neighbours is diagonal in momentum space. The crystal momentum $\mathbf{k}$ is a good quantum number labelling each eigenstate. On the other hand, the SL corresponds to a tight-binding model in momentum space which has good quantum numbers $\mathbf{r}$ in position space. The dynamics of $\mathbf{r}$ in an SL is analogous to the dynamics of $\mathbf{k}$ in a real space lattice. We show that Bloch oscillations, Wannier Stark ladders and Bloch band collapsing can be observed in an SL based on electromagnetically induced transparency (EIT). The two-dimensional SL provides a tunable quantum optics model for Dirac physics in graphene.

The momentum transfer between a single two-level atom and standing wave light is quantized. The states of the atom with quantized recoil momenta thus have discrete translational symmetry in momentum space \cite{Cooper2013}. To inhibit the recoil motions, we can use fixed three-level systems in solids, which effectively have infinite mass thanks to the Lamb-M\"ossbauer effect \cite{Lamb1939, Mossbauer1958,Frauenfelder1962}. The phase correlations of the timed Dicke states, rather than the recoil momenta of single atoms, set the lattice points in momentum space. 

\begin{figure}[t]
    \epsfig{figure=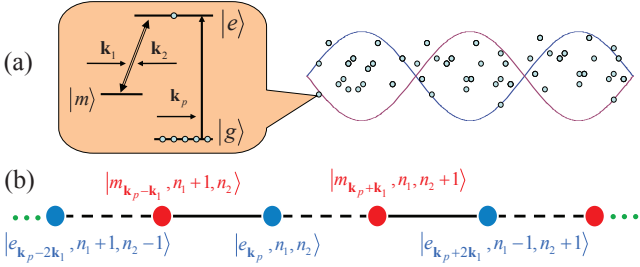, angle=0, width=0.5\textwidth}
\caption{(Color online) (a) The real space configuration and the internal atomic states of a 1D bipartite SL in momentum space. An EM plane wave mode $\mathbf{k}_p$ collectively excites the transition from $|g\ra$ to $|e\ra$. The standing wave formed by modes $\mathbf{k}_1$ and $\mathbf{k}_2$ couples the transition between $|e\ra$ and $|m\ra$. (b) The 1D bipartite SL in momentum space. The red (blue) circles represent the $|m_{\mathbf{k}}\ra$ ($|e_{\mathbf{k}}\ra$) states. The solid (dashed) lines represent the interaction via mode $\mathbf{k}_1$ ($\mathbf{k}_2$). The distance between the adjacent sites is $|\mathbf{k}_1|$ and the direction of $\mathbf{k}_1$ is defined to the right.}
\label{01d}
\end{figure}

\emph{Dicke spinor}.---A collection of $N$ two-level atoms coupled by a single electromagnetic (EM) mode is described by the Dicke model \cite{Dicke1954}. If the atoms are randomly distributed in an area much larger than the wavelength, the first excited state which is the timed Dicke state \cite{Scully2006} can record the momentum of the absorbed photon via phase correlations between excited atoms,
\begin{equation}
|e_{\mathbf{k}_p}\ra=\frac{1}{\sqrt{N}}\sum\limits_{\alpha=1}^{N}e^{i\mathbf{k}_p\cdot\mathbf{r}_\alpha}|g_1, g_2,...,e_\alpha,...,g_N\ra.
\label{toD}
\end{equation}
Here $\mathbf{k}_p$ is the wave vector of the photon, $\mathbf{r}_\alpha$ is the position of the $\alpha$th atom, which has the ground state $|g_\alpha\ra$ and excited state $|e_\alpha\ra$. The atomic levels are shown in the inset of Fig.\ref{01d} (a). Now we apply another EM plane wave mode $\mathbf{k}_1$ that couples $|e\ra$ to a metastable state $|m\ra$ via the interaction Hamiltonian (in the rotating wave approximation)
$H_I=-\sum_{\alpha=1}^{N}\hbar \kappa_1 a_1 e^{i\mathbf{k}_1\cdot\mathbf{r}_\alpha} |e_\alpha\ra\la m_\alpha|+h.c.$,
where $a_1$ is the annihilation operator of mode $\mathbf{k}_1$, $\kappa_1$ is the vacuum coupling strength (assumed to be real). Then $|e_{\mathbf{k}_p}, n_1\ra$ is coupled to $|m_{\mathbf{k}_p-\mathbf{k}_1}, n_1+1\ra$  where $n_1$ is the photon number of mode $\mathbf{k}_1$, and $|m_{\mathbf{k}_p-\mathbf{k}_1}\ra$ is defined by replacing $e$ with $m$ in Eq. (\ref{toD}). The Rabi frequency $\kappa_1 \sqrt{n_1+1}$ is independent of the atom number $N$. The two-states $|e_{\mathbf{k}_p}, n_1\ra$ and $|m_{\mathbf{k}_p-\mathbf{k}_1}, n_1+1\ra$ form a two-component Dicke spinor.

\emph{1D bipartite SL}.---By introducing a second mode $\mathbf{k}_2=-\mathbf{k}_1$, the interaction Hamiltonian
\begin{equation}
H_I=-\sum\limits_{\alpha=1}^{N}\hbar (\kappa_1 a_1 e^{i\mathbf{k}_1\cdot\mathbf{r}_\alpha}+\kappa_2 a_2 e^{i\mathbf{k}_2\cdot\mathbf{r}_\alpha}) |e_\alpha\ra\la m_\alpha|+h.c.,
\label{H1}
\end{equation}
extends the Dicke spinor to a one-dimensional (1D) bipartite SL, as shown in Fig.\ref{01d} (a). The state $|e_{\mathbf{k}_p}, n_1, n_2\ra$ can be coupled either by mode $\mathbf{k}_1$ to $|m_{\mathbf{k}_p-\mathbf{k}_1}, n_1+1, n_2\ra$, or by mode $\mathbf{k}_2$ to $|m_{\mathbf{k}_p+\mathbf{k}_1}, n_1, n_2+1\ra$, as shown in Fig.\ref{01d} (b). The Rabi frequencies are site-dependent. However, if the two fields are in coherent states with large average photon numbers $\la n_i\ra\gg 1$ $(i=1,2)$, the Rabi frequencies are approximately constant $\Omega_i=\kappa_i\sqrt{\la n_i\ra}$. We can rewrite the interaction Hamiltonian in Eq. (\ref{H1}) into a tight-binding form,
\begin{equation}
H_I=-\sum\limits_{j} (\hbar \Omega_1 \hat{e}^\dagger_{j} \hat{m}_{j}+
\hbar \Omega_2 \hat{m}^\dagger_{j+1} \hat{e}_{j})+h.c.,
\label{H1k}
\end{equation}
where the creation operators in the $j$th unit cell are
\begin{equation}
         \begin{aligned}
           \hat{e}^\dagger_j&=\frac{1}{\sqrt{N}}\sum\limits_{\alpha=1}^{N} e^{i(\mathbf{k}_p+2j\mathbf{k}_1)\cdot\mathbf{r}_\alpha} |e_\alpha\ra \la g_\alpha|,\\
           \hat{m}^\dagger_j&=\frac{1}{\sqrt{N}}\sum_{\alpha=1}^{N} e^{i[\mathbf{k}_p+(2j-1)\mathbf{k}_1]\cdot\mathbf{r}_\alpha} |m_\alpha\ra \la g_\alpha|,
          \end{aligned}
\end{equation}
with $|G\ra\equiv|g_1, g_2,..., g_N\ra$ the ground state and the superradiant states $|e_{\mathbf{k}_p+2j\mathbf{k}_1}\ra=\hat{e}^\dagger_j|G\ra$. The Hamiltonian in Eq.(\ref{H1k}) is also valid for many excitations if the excitation number is much less than the atom number, and the operators are approximately bosonic, $[\hat{e}_j,\hat{e}^\dagger_{j^\prime}]=\delta_{jj^\prime}$.

\begin{figure}[t]
    \epsfig{figure=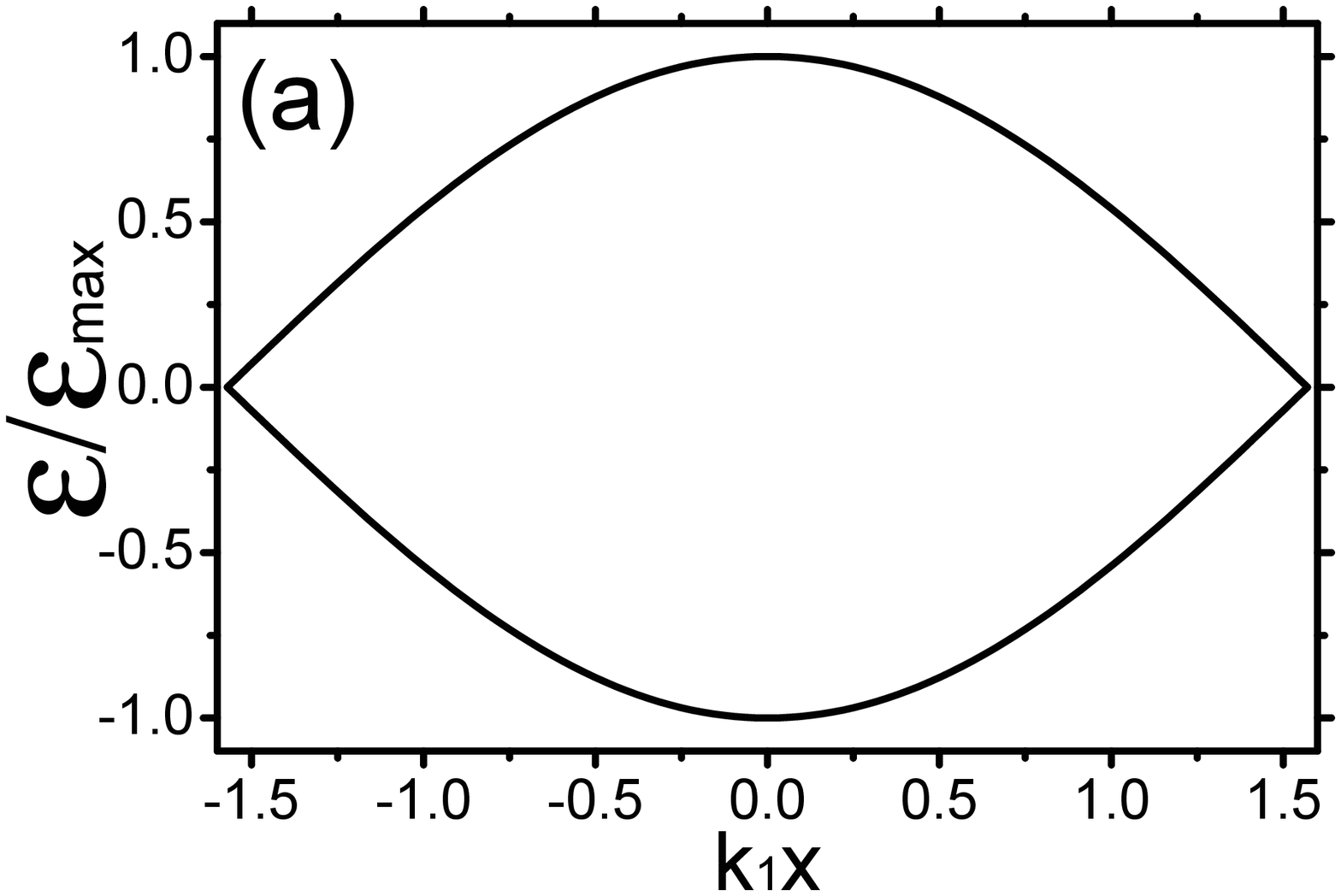, angle=0, width=0.21\textwidth}
    \epsfig{figure=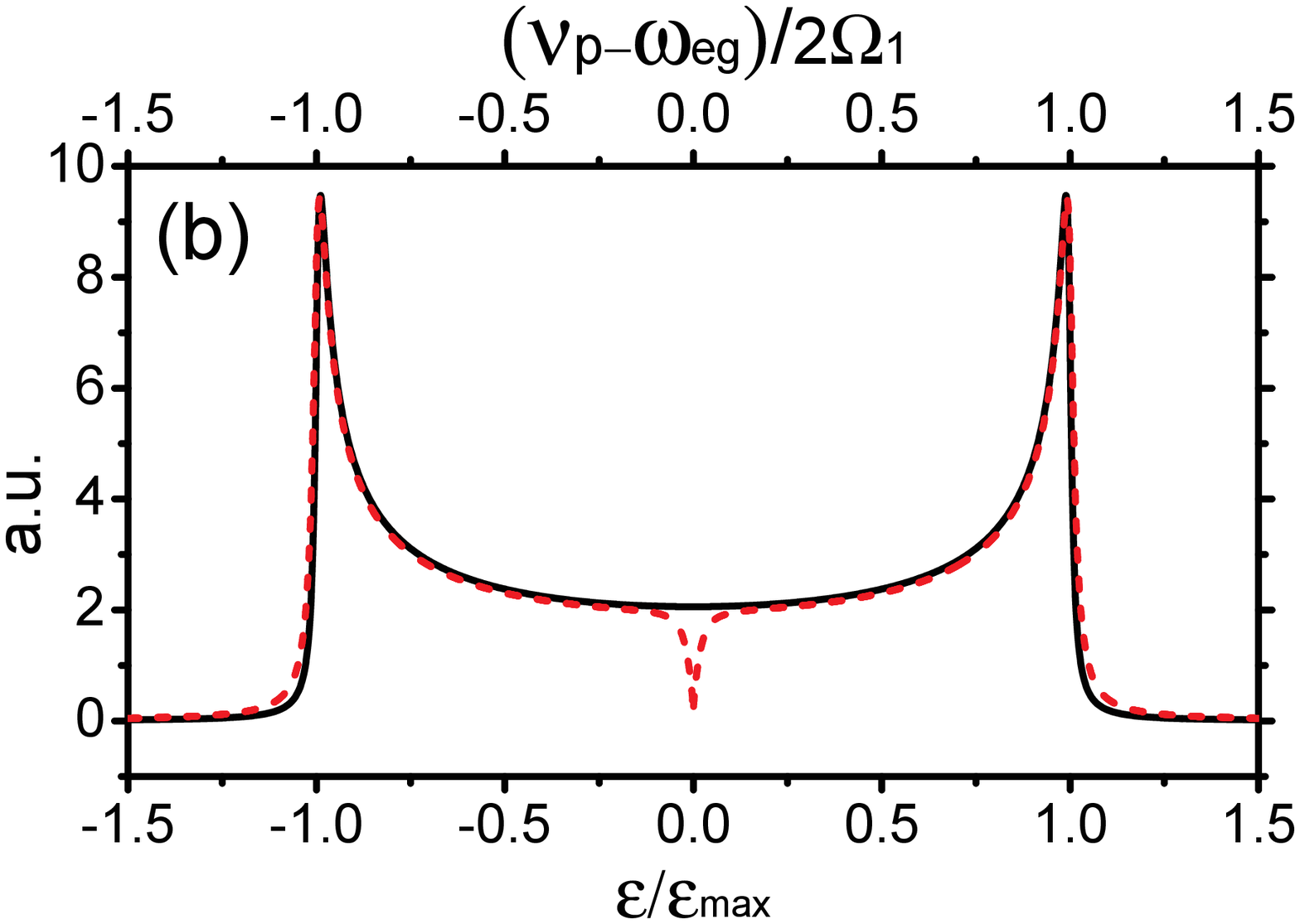, angle=0, width=0.24\textwidth}
\caption{(Color online) (a) The dispersion relation of a 1D SL. (b) The DOS of the SL (black solid) and the standing wave coupled EIT absorption spectrum (red dash). $\gamma=0.06\Omega_1$, $\gamma^\prime=0$. Assuming that each eigenstate has a finite life time, the DOS is Lorentzian broadened with width 0.01$\epsilon_{\text{max}}$ to fit with the EIT absorption spectrum.}
\label{da}
\end{figure}

The tight-binding model in momentum space is diagonal in its reciprocal position space. For simplicity, we let $\Omega_1=\Omega_2$. The dispersion relation is
\begin{equation}
\epsilon_{\pm}\left(\mathbf{r}\right)=\pm 2\hbar \Omega_1 \cos( \mathbf{r}\cdot\mathbf{k}_1),
\label{dispersion}
\end{equation}
as shown in Fig.\ref{da} (a). The energy band is directly shown by the interference pattern of the coupling standing wave.

\emph{Detection by the standing wave coupled EIT}.---Levels $|e\ra$ and $|m\ra$ are resonantly coupled by EM modes $\mathbf{k}_1=k_1\hat{x}$ and $\mathbf{k}_2=-k_1\hat{x}$. A weak field $\mathbf{k}_p=k_p\hat{x}$ which probes the transition from the ground state $|g\ra$ to level $|e\ra$ should create excitations in the 1D SL. The density of states (DOS) of the SL, $D(\epsilon)=N/\pi\sqrt{\epsilon^2_{\text{max}}-\epsilon^2}$ with $\epsilon_{\text{max}}=2\hbar\Omega_1$, can therefore be tested by the absorption spectrum of $\mathbf{k}_p$, which on the other hand can be obtained from the imaginary part of the EIT susceptibility \cite{Artoni2006,Supplementary},
\begin{equation}
\chi\left( x \right)
=\frac{3\pi \mathcal{N}{\Gamma}(\Delta_p-i\gamma^\prime)}{({\Delta }_{p}-i\gamma^\prime)({\Delta }_{p}-i{\gamma})
-{{\left| {{\Omega }_{1}}{{e}^{i{{k}_{1}}x}}+{{\Omega}_{2}}{{e}^{-i{{k}_{1}}x}} \right|}^{2}}}.
\label{xi}
\end{equation}
Here $\mathcal{N}$ is the atomic numbers in the volume $c^{3}/\omega^3_{eg}$ where $\omega_{eg}$ is the transition frequency between $|e\ra$ and $|g\ra$. $\gamma$ and $\Gamma$ are the decoherence rate and radiative decay rate between $|e\ra$ and $|g\ra$, respectively. $\gamma^\prime$ is the decoherence rate between $|g\ra$ and $|m\ra$. $\Delta_p=\omega_{eg}-\nu_p$ is the detuning of the probe field.

\begin{figure}[t]
    \epsfig{figure=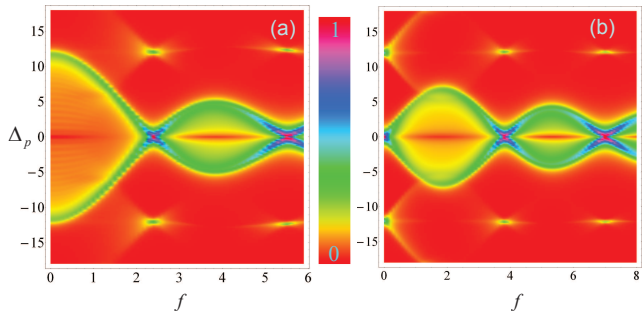, angle=0, width=0.5\textwidth}
  \caption{(Color online) The absorption spectra of 1D SL in an oscillating force. (a) Absorption spectra as function of detuning $\Delta_p$ and the amplitude-frequency ratio of the effective oscillating force $f$. $\delta_1=\delta_2=0$. The band collapses at the Bessel function zeros $J_0(f)=0$ for $f=2.4$ and $5.5$. (b) Absorption spectra for $\delta_1=-\delta_2=\nu_d$. The band collapses at the Bessel function zeros $J_1(f)=0$ for $f=3.8$ and $7.0$. The other parameters are $\Omega_1=\Omega_2=6\gamma$, $\nu_d=12\gamma$, $\gamma=1$, and $\gamma'=0.001$.} 
  \label{bgclps} 
\end{figure}

The absorption in Eq. (\ref{xi}) is periodic in space. The total absorption spectrum can be obtained by averaging Eq. (\ref{xi}) over one period,
\begin{equation}
A\left(\nu_p\right)\propto\text{Im}\left[\frac{k_1}{\pi}\int_{-\frac{\pi}{2k_1}}^{\frac{\pi}{2k_1}}\chi(x)\text{d} x\right].
\end{equation}
In Fig.\ref{da} (b), we plotted the density of states $D$ and the absorption spectrum $A$. Their overlap demonstrates the equivalence between the 1D SL and the standing wave coupled EIT. The major difference is that the absorption spectrum $A$ has a transparency point at zero detuning due to the Fano interference \cite{Fano1961, Miroshnichenko2010}. 

\emph{Effective force in momentum space}.---An effective uniform force in an SL should introduce a potential linear with the momenta of the sites. From Fig.\ref{01d} (b), we see that the superradiant states are correlated with photon numbers that are linear with the momenta. We therefore can introduce an effective uniform force in momentum space by changing the energy of the photons of the two modes. The unperturbed Hamiltonian is
\begin{equation}
\begin{aligned}
H_0=&\sum\limits_{\alpha=1}^{N}\hbar \left(\omega_e |e_\alpha\ra\la e_\alpha|+ \omega_m |m_\alpha\ra\la m_\alpha|\right)\\
&+\hbar \nu_1 a^\dagger_1 a_1+\hbar \nu_2 a^\dagger_2 a_2,
\end{aligned}
\end{equation}
where $\hbar\omega_i$ $(i=e,m)$ is the atomic eigenenergy and $\nu_i$ $(i=1,2)$ is the angular frequency of the fields. The energy difference between $|e_{\mathbf{k}_p}, n_1, n_2\ra$ and $|m_{\mathbf{k}_p-\mathbf{k}_1}, n_1+1, n_2\ra$ is $\hbar\delta_1=\hbar\omega_{em}-\hbar\nu_1$ where $\omega_{em}=\omega_e-\omega_m$, and the energy difference between $|e_{\mathbf{k}_p}, n_1, n_2\ra$ and $|m_{\mathbf{k}_p+\mathbf{k}_1}, n_1, n_2+1\ra$ is $\hbar\delta_2=\hbar\omega_{em}-\hbar\nu_2$. The quantity $\hbar\delta_0=\hbar \omega_{em}-\frac{1}{2}(\nu_1+\nu_2)$ is the energy difference between the two sublattices of $|e\ra$ and $|m\ra$. The detuning between the two fields $2\delta=\nu_1-\nu_2$ is the potential difference between adjacent unit cells. The potential is linear of the momentum $\mathbf{p}=\hbar\mathbf{k}=-i\hbar\sum_{\alpha}\nabla_{\mathbf{r}_\alpha}$,
\begin{equation}
V(\mathbf{p})=-\boldsymbol{\mathcal{F}}\cdot\mathbf{p},
\label{elec}
\end{equation}
where the momentum-space force $\boldsymbol{\mathcal{F}}=-\nabla_{\mathbf{p}}V(\mathbf{p})=\frac{\delta}{k_1}\hat{\mathbf{k}}_1$ is in contrast to the real-space force \cite{Holthaus1992, Wilkinson1996, Madison1998, Szameit2010,Supplementary}. Therefore, the effective Hamiltonian is
\begin{equation}
\begin{aligned}
H
&=\sum\limits_{j} \hbar(\delta_0-2j\delta)\hat{e}^\dagger_{j}\hat{e}_{j}-\hbar(2j-1)\delta\hat{m}^\dagger_{j}\hat{m}_{j}\\
&-\left(\hbar \Omega_1 \hat{e}^\dagger_{j} \hat{m}_{j}+
\hbar \Omega_2 \hat{m}^\dagger_{j+1} \hat{e}_{j}+h.c.\right).
\end{aligned}
\end{equation}
The equation of motion of the position operator $\mathbf{r}_\alpha$ of the $\alpha$th atom is
\begin{equation}
\dot{\mathbf{r}}_\alpha=\frac{1}{i\hbar}\left[\mathbf{r}_\alpha, -\boldsymbol{\mathcal{F}}\cdot\mathbf{p}\right]=-\boldsymbol{\mathcal{F}}=-\frac{\delta}{k_1}\hat{\mathbf{k}}_1.
\label{vel}
\end{equation}
It is easy to understand this equation in real space. The detuning $\delta$ leads to a moving standing wave with velocity $-\dot{\mathbf{r}}_\alpha$. By adiabatic following, the position $\mathbf{r}_\alpha$ will move with the velocity $\dot{\mathbf{r}}_\alpha$. After time $T=\pi/\delta$, the standing wave moves a period $\lambda_1/2=\pi/k_1$ and the system recovers its original state, which is the Bloch oscillation in the SL.

\emph{Bloch band collapsing}.---If the effective force $\boldsymbol{\mathcal{F}}=\mathcal{F}(t)\hat{x}$ is periodic in time, the band collapsing may occur \cite{Dunlap1986, Holthaus1992, Zak1993}. We make the frequencies of the two fields time-dependent, $\nu_i+\Delta_i\cos\nu_d t$ $(i=1,2)$, which introduces an oscillating force in the SL. In particular for $\delta_1=-\delta_2=n\nu_d$ with integer $n$ and $\Delta_2=-\Delta_1$, the excitation in the SL is driven by an effective force $\mathcal{F}(t)=\mathcal{F}_s+\mathcal{F}_d\cos\nu_d t$ with static component $\mathcal{F}_s=-n\nu_d/k_1$ and dynamic component $\mathcal{F}_d=\Delta_1/k_1$. The quasienergy band is \cite{Zak1993}
\begin{equation}
\epsilon_n\left(x\right)=\pm 2\Omega_1 J_n\left(f\right)\cos (x k_1),
\label{Jn}
\end{equation}
where $J_n(f)$ is the $n$th order Bessel function of the first kind and $f=\Delta_1/\nu_d$. One interesting feature of this Floquet quasienergy band is that it collapses at the zeros of $J_n(f)$. 

Fig.\ref{bgclps} (a) shows the EIT absorption spectra associated with the quasienergy bands for $n=0$ \cite{Supplementary}. At $f=0$, the absorption spectrum has a broad DOS of a bipartite lattice. Increasing $f$ leads to a narrower energy band following $J_0(f)$ and finally the energy band collapses at $f=2.4$, where a strong absorption peak appears at the zero detuning. The separation between the Floquet energy bands $\nu_d=2\Omega_1$ is large and the interaction between the states from different bands is weak. Therefore, most of the upper and lower Floquet bands are not visible. However, near the band collapsing points $f=2.4$, the two Floquet sidebands is vaguely visible due to the large DOS.

The Wannier-Stark ladder appears if the force has a static part. In Fig.\ref{bgclps} (b), we plot the absorption spectra for $n=1$. If $f=0$, the force is purely static, and three peaks are shown at $\Delta_p=0$, $\pm \nu_d$, which are the energies of the states in the Wannier-Stark ladder. As $f$ increases, we observe energy bands following Eq. (\ref{Jn}) with $n=1$. The bands collapse at the zero points of $J_1(f)$, $f=3.8$ and $7.0$. These are consistent with the results of electrons \cite{Zhao1991, Holthaus1995, Liu1999}. The band collapsing for some other cases are discussed in the Supplementary Material \cite{Supplementary}.

\begin{figure}[t]
    \epsfig{figure=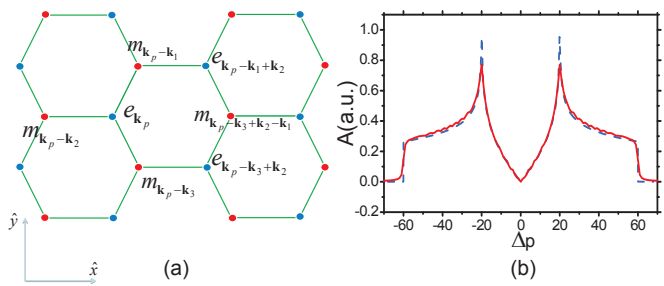, angle=0, width=0.5\textwidth}
  \caption{(Color online) (a) The graphene structure of the 2D SL. Timed Dicke states with $|e\ra$ and $|m\ra$ correspond to the two sublattices of graphene. The three bonds to the nearest neighbour correspond to the three coupling fields. (b) The DOS of graphene (blue dash) and the absorption spectrum of the 2D SL (red solid). The Rabi frequencies of the three coupling fields are all $20\gamma$. $\gamma=1$, $\gamma^\prime=0.001$. Correspondingly, the nearest-neighbour hopping coefficents of graphene are set as 20.} 
  \label{2dsl} 
\end{figure}

\emph{2D and higher dimensional SL}.---The 1D superradiance chain can be extended to a 2D honeycomb lattice by introducing three-mode coupling field with wave vectors $\mathbf{k}_1=k(-\frac{1}{2}\hat{x}-\frac{\sqrt{3}}{2}\hat{y})$, $\mathbf{k}_2=k\hat{x}$ and $\mathbf{k}_3=k(-\frac{1}{2}\hat{x}+\frac{\sqrt{3}}{2}\hat{y})$, as shown in Fig.\ref{2dsl} (a). In Fig.\ref{2dsl} (b), we plot the 2D SL absorption spectrum, which overlaps with the DOS of graphene \cite{Novoselov2005,CastroNeto2009}. A striking feature is that the EIT point in the 2D SL at $\Delta_p=0$ corresponds to the Dirac point in graphene \cite{Haldane1988}. The 2D SL provides a highly tunable platform for the Dirac physics in graphene \cite{Novoselov2005}, whose material properties are fixed. The hopping coefficients and on-site potentials can be easily tuned by the strengths and frequencies of the coupling fields. Interesting graphene physics in the 2D SL, such as Berry phases, artificial gauge field and Haldane model \cite{Haldane1988} will be discussed elsewhere \cite{Wang2015}.

Similarly, four-mode coupling field can construct diamond-structure tight-binding models. A particular interesting subject to be investigated in the future is the tight-binding SL in dimensions higher than three when the number of the coupling fields is more than four. Since real-space tight-binding models have at most three-dimensions, a wealth of new physics such as the 4D quantum Hall effect \cite{Zhang2001} may emerge from the extra dimensions of SL.

\emph{Discussion}.---The quantum dynamics of the 1D SL can be detected by the electromagnetically induced grating (EIG) \cite{Ling1998} where the $n$th order diffraction is emitted by the superradiant state $|e_{\mathbf{k}_p+2n\mathbf{k}_1}\ra$. EIG \cite{Mitsunaga1999, Cardoso2002} and bi-chromatic EIT \cite{Wang2003} have been experimentally observed without being related to tight-binding lattices. Since they only focused on the dynamics of the light, the rich physics concerning the timed Dicke states (many of them are subradiant and thus not detectable in the diffraction of EIG) were ignored. Nevertheless, EIG and the related experiments can be further used to observe the dynamic localization via the disappearance of the EIG diffraction, and many other condensed-matter phenomena in the 2D and higher dimensional SL. 

The SL can be realized in cold atoms if the Doppler shift due to the recoil is much smaller than the coupling field Rabi frequency \cite{Supplementary}. For example, we can choose $^{85}$Rb D1 line with $|g\ra=|5^{2}S_{1/2}, F=2\ra$, $|e\ra=|5^{2}P_{1/2}, F=2\ra$ and $|m\ra=|5^{2}S_{1/2}, F=3\ra$. The decoherence rate $\gamma=2.9$MHz. For the parameters in Fig.\ref{bgclps}, the Rabi frequency $\Omega_1=6\gamma=17.3$MHz (intensity 0.1W/cm$^{2}$) is much larger than the recoil Doppler shift $7.4$kHz. The modulation frequency $\nu_d=12\gamma=34.5$MHz and the modulation amplitude $\Delta_{1,2}$ should be in the range $0\sim 200$MHz. In the $\mu$K regime, the thermal motions induce frequency shifts $\sim$kHz, which are much smaller than the effective potential energy $\sim$MHz. One can easily trap $10^{6}$ atoms in $1$ mm$^3$ and a Gaussian beam with 3 mm diameter can be approximated as a plane wave \cite{Wang2003}. The band collapsing can be directly observed through the absorption spectra.

The applications of SL's are promising. The transport of the superradiant excitations in SL's can be used to reflect high-frequency light (for example, x-ray or ultraviolet) with low-frequency light (visible light or infrared) \cite{2013arXiv1305.3636W}. The coupling strength between the lattice point is tunable, which allows us to prepare a superposition of two timed Dicke states that are far apart in momentum space for Heisenberg limit metrology \cite{Wang2014}. The effective force in momentum space can break the time-reversal symmetry and realize optical isolation \cite{Wang2013}.

In conclusion, we proposed the concept of superradiance lattices based on a standing wave coupled EIT system. An effective uniform force in momentum space can be introduced by the detuning between the two components of the standing wave. The Wannier-Stark ladder and the Bloch band collapsing can be observed from the absorption spectra of the probe field. The dynamic localization can be observed from the disappearance of diffraction in an EIG scheme. By introducing more EM modes, this lattice can be extended to higher dimensions where many interesting physics can be studied.

We thank R. Glauber, W. P. Schleich and J. Evers for helpful discussion. We gratefully acknowledge the support of the National Science Foundation Grants No. PHY-1241032 (INSPIRE CREATIV) and PHY-1068554 and the Robert A. Welch Foundation (Grant No. A-1261). R.-B.Liu was supported by Hong Kong RGC Project 401011 and CUHK Focused Investments Scheme. S.-Y. Zhu was supported by NSFC1174026.

\bibliographystyle{apsrev4-1}

\begin{thebibliography}{65}%
\makeatletter
\providecommand \@ifxundefined [1]{%
 \@ifx{#1\undefined}
}%
\providecommand \@ifnum [1]{%
 \ifnum #1\expandafter \@firstoftwo
 \else \expandafter \@secondoftwo
 \fi
}%
\providecommand \@ifx [1]{%
 \ifx #1\expandafter \@firstoftwo
 \else \expandafter \@secondoftwo
 \fi
}%
\providecommand \natexlab [1]{#1}%
\providecommand \enquote  [1]{``#1''}%
\providecommand \bibnamefont  [1]{#1}%
\providecommand \bibfnamefont [1]{#1}%
\providecommand \citenamefont [1]{#1}%
\providecommand \href@noop [0]{\@secondoftwo}%
\providecommand \href [0]{\begingroup \@sanitize@url \@href}%
\providecommand \@href[1]{\@@startlink{#1}\@@href}%
\providecommand \@@href[1]{\endgroup#1\@@endlink}%
\providecommand \@sanitize@url [0]{\catcode `\\12\catcode `\$12\catcode
  `\&12\catcode `\#12\catcode `\^12\catcode `\_12\catcode `\%12\relax}%
\providecommand \@@startlink[1]{}%
\providecommand \@@endlink[0]{}%
\providecommand \url  [0]{\begingroup\@sanitize@url \@url }%
\providecommand \@url [1]{\endgroup\@href {#1}{\urlprefix }}%
\providecommand \urlprefix  [0]{URL }%
\providecommand \Eprint [0]{\href }%
\providecommand \doibase [0]{http://dx.doi.org/}%
\providecommand \selectlanguage [0]{\@gobble}%
\providecommand \bibinfo  [0]{\@secondoftwo}%
\providecommand \bibfield  [0]{\@secondoftwo}%
\providecommand \translation [1]{[#1]}%
\providecommand \BibitemOpen [0]{}%
\providecommand \bibitemStop [0]{}%
\providecommand \bibitemNoStop [0]{.\EOS\space}%
\providecommand \EOS [0]{\spacefactor3000\relax}%
\providecommand \BibitemShut  [1]{\csname bibitem#1\endcsname}%
\let\auto@bib@innerbib\@empty
\bibitem [{\citenamefont {Bloch}(1929)}]{Bloch1929}%
  \BibitemOpen
  \bibfield  {author} {\bibinfo {author} {\bibfnamefont {F.}~\bibnamefont
  {Bloch}},\ }\href@noop {} {\bibfield  {journal} {\bibinfo  {journal}
  {Zeitschrift f\"ur Physik}\ }\textbf {\bibinfo {volume} {52}},\ \bibinfo
  {pages} {555} (\bibinfo {year} {1929})}\BibitemShut {NoStop}%
\bibitem [{\citenamefont {Zener}(1934)}]{Zener1934}%
  \BibitemOpen
  \bibfield  {author} {\bibinfo {author} {\bibfnamefont {C.}~\bibnamefont
  {Zener}},\ }\href@noop {} {\bibfield  {journal} {\bibinfo  {journal}
  {Proceedings of the Royal Society of London. Series A}\ }\textbf {\bibinfo
  {volume} {145}},\ \bibinfo {pages} {523} (\bibinfo {year}
  {1934})}\BibitemShut {NoStop}%
\bibitem [{\citenamefont {Wannier}(1960)}]{Wannier1960}%
  \BibitemOpen
  \bibfield  {author} {\bibinfo {author} {\bibfnamefont {G.~H.}\ \bibnamefont
  {Wannier}},\ }\href@noop {} {\bibfield  {journal} {\bibinfo  {journal}
  {Physical Review}\ }\textbf {\bibinfo {volume} {117}},\ \bibinfo {pages}
  {432} (\bibinfo {year} {1960})}\BibitemShut {NoStop}%
\bibitem [{\citenamefont {Ignatov}\ and\ \citenamefont
  {Romanov}(1976)}]{Ignatov1976}%
  \BibitemOpen
  \bibfield  {author} {\bibinfo {author} {\bibfnamefont {A.~A.}\ \bibnamefont
  {Ignatov}}\ and\ \bibinfo {author} {\bibfnamefont {Y.~A.}\ \bibnamefont
  {Romanov}},\ }\href@noop {} {\bibfield  {journal} {\bibinfo  {journal}
  {Physica Status Solidi B-Basic Research}\ }\textbf {\bibinfo {volume} {73}},\
  \bibinfo {pages} {327} (\bibinfo {year} {1976})}\BibitemShut {NoStop}%
\bibitem [{\citenamefont {Dunlap}\ and\ \citenamefont
  {Kenkre}(1986)}]{Dunlap1986}%
  \BibitemOpen
  \bibfield  {author} {\bibinfo {author} {\bibfnamefont {D.~H.}\ \bibnamefont
  {Dunlap}}\ and\ \bibinfo {author} {\bibfnamefont {V.~M.}\ \bibnamefont
  {Kenkre}},\ }\href@noop {} {\bibfield  {journal} {\bibinfo  {journal}
  {Physical Review B}\ }\textbf {\bibinfo {volume} {34}},\ \bibinfo {pages}
  {3625} (\bibinfo {year} {1986})}\BibitemShut {NoStop}%
\bibitem [{\citenamefont {Leo}\ \emph {et~al.}(1992)\citenamefont {Leo},
  \citenamefont {Bolivar}, \citenamefont {Br\"uggemann}, \citenamefont
  {Schwedler},\ and\ \citenamefont {K\"ohler}}]{Leo1992}%
  \BibitemOpen
  \bibfield  {author} {\bibinfo {author} {\bibfnamefont {K.}~\bibnamefont
  {Leo}}, \bibinfo {author} {\bibfnamefont {P.~H.}\ \bibnamefont {Bolivar}},
  \bibinfo {author} {\bibfnamefont {F.}~\bibnamefont {Br\"uggemann}}, \bibinfo
  {author} {\bibfnamefont {R.}~\bibnamefont {Schwedler}}, \ and\ \bibinfo
  {author} {\bibfnamefont {K.}~\bibnamefont {K\"ohler}},\ }\href@noop {}
  {\bibfield  {journal} {\bibinfo  {journal} {Solid State Communications}\
  }\textbf {\bibinfo {volume} {84}},\ \bibinfo {pages} {943} (\bibinfo {year}
  {1992})}\BibitemShut {NoStop}%
\bibitem [{\citenamefont {Lyssenko}\ \emph {et~al.}(1997)\citenamefont
  {Lyssenko}, \citenamefont {Valu\v{s}is}, \citenamefont {L\"oser},
  \citenamefont {Hasche}, \citenamefont {Leo}, \citenamefont {Dignam},\ and\
  \citenamefont {K\"ohler}}]{Lyssenko1997}%
  \BibitemOpen
  \bibfield  {author} {\bibinfo {author} {\bibfnamefont {V.~G.}\ \bibnamefont
  {Lyssenko}}, \bibinfo {author} {\bibfnamefont {G.}~\bibnamefont
  {Valu\v{s}is}}, \bibinfo {author} {\bibfnamefont {F.}~\bibnamefont
  {L\"oser}}, \bibinfo {author} {\bibfnamefont {T.}~\bibnamefont {Hasche}},
  \bibinfo {author} {\bibfnamefont {K.}~\bibnamefont {Leo}}, \bibinfo {author}
  {\bibfnamefont {M.~M.}\ \bibnamefont {Dignam}}, \ and\ \bibinfo {author}
  {\bibfnamefont {K.}~\bibnamefont {K\"ohler}},\ }\href@noop {} {\bibfield
  {journal} {\bibinfo  {journal} {Physical Review Letters}\ }\textbf {\bibinfo
  {volume} {79}},\ \bibinfo {pages} {301} (\bibinfo {year} {1997})}\BibitemShut
  {NoStop}%
\bibitem [{\citenamefont {Ben~Dahan}\ \emph {et~al.}(1996)\citenamefont
  {Ben~Dahan}, \citenamefont {Peik}, \citenamefont {Reichel}, \citenamefont
  {Castin},\ and\ \citenamefont {Salomon}}]{BenDahan1996}%
  \BibitemOpen
  \bibfield  {author} {\bibinfo {author} {\bibfnamefont {M.}~\bibnamefont
  {Ben~Dahan}}, \bibinfo {author} {\bibfnamefont {E.}~\bibnamefont {Peik}},
  \bibinfo {author} {\bibfnamefont {J.}~\bibnamefont {Reichel}}, \bibinfo
  {author} {\bibfnamefont {Y.}~\bibnamefont {Castin}}, \ and\ \bibinfo {author}
  {\bibfnamefont {C.}~\bibnamefont {Salomon}},\ }\href@noop {} {\bibfield
  {journal} {\bibinfo  {journal} {Physical Review Letters}\ }\textbf {\bibinfo
  {volume} {76}},\ \bibinfo {pages} {4508} (\bibinfo {year}
  {1996})}\BibitemShut {NoStop}%
\bibitem [{\citenamefont {Wilkinson}\ \emph {et~al.}(1996)\citenamefont
  {Wilkinson}, \citenamefont {Bharucha}, \citenamefont {Madison}, \citenamefont
  {Niu},\ and\ \citenamefont {Raizen}}]{Wilkinson1996}%
  \BibitemOpen
  \bibfield  {author} {\bibinfo {author} {\bibfnamefont {S.~R.}\ \bibnamefont
  {Wilkinson}}, \bibinfo {author} {\bibfnamefont {C.~F.}\ \bibnamefont
  {Bharucha}}, \bibinfo {author} {\bibfnamefont {K.~W.}\ \bibnamefont
  {Madison}}, \bibinfo {author} {\bibfnamefont {Q.}~\bibnamefont {Niu}}, \ and\
  \bibinfo {author} {\bibfnamefont {M.~G.}\ \bibnamefont {Raizen}},\
  }\href@noop {} {\bibfield  {journal} {\bibinfo  {journal} {Physical Review
  Letters}\ }\textbf {\bibinfo {volume} {76}},\ \bibinfo {pages} {4512}
  (\bibinfo {year} {1996})}\BibitemShut {NoStop}%
\bibitem [{\citenamefont {Holthaus}(1992)}]{Holthaus1992}%
  \BibitemOpen
  \bibfield  {author} {\bibinfo {author} {\bibfnamefont {M.}~\bibnamefont
  {Holthaus}},\ }\href@noop {} {\bibfield  {journal} {\bibinfo  {journal}
  {Physical Review Letters}\ }\textbf {\bibinfo {volume} {69}},\ \bibinfo
  {pages} {351} (\bibinfo {year} {1992})}\BibitemShut {NoStop}%
\bibitem [{\citenamefont {Graham}\ \emph {et~al.}(1992)\citenamefont {Graham},
  \citenamefont {Schlautmann},\ and\ \citenamefont {Zoller}}]{Graham1992}%
  \BibitemOpen
  \bibfield  {author} {\bibinfo {author} {\bibfnamefont {R.}~\bibnamefont
  {Graham}}, \bibinfo {author} {\bibfnamefont {M.}~\bibnamefont {Schlautmann}},
  \ and\ \bibinfo {author} {\bibfnamefont {P.}~\bibnamefont {Zoller}},\
  }\href@noop {} {\bibfield  {journal} {\bibinfo  {journal} {Physical Review
  A}\ }\textbf {\bibinfo {volume} {45}},\ \bibinfo {pages} {R19} (\bibinfo
  {year} {1992})}\BibitemShut {NoStop}%
\bibitem [{\citenamefont {Moore}\ \emph {et~al.}(1994)\citenamefont {Moore},
  \citenamefont {Robinson}, \citenamefont {Bharucha}, \citenamefont
  {Williams},\ and\ \citenamefont {Raizen}}]{Moore1994}%
  \BibitemOpen
  \bibfield  {author} {\bibinfo {author} {\bibfnamefont {F.~L.}\ \bibnamefont
  {Moore}}, \bibinfo {author} {\bibfnamefont {J.~C.}\ \bibnamefont {Robinson}},
  \bibinfo {author} {\bibfnamefont {C.}~\bibnamefont {Bharucha}}, \bibinfo
  {author} {\bibfnamefont {P.~E.}\ \bibnamefont {Williams}}, \ and\ \bibinfo
  {author} {\bibfnamefont {M.~G.}\ \bibnamefont {Raizen}},\ }\href@noop {}
  {\bibfield  {journal} {\bibinfo  {journal} {Physical Review Letters}\
  }\textbf {\bibinfo {volume} {73}},\ \bibinfo {pages} {2974} (\bibinfo {year}
  {1994})}\BibitemShut {NoStop}%
\bibitem [{\citenamefont {Madison}\ \emph {et~al.}(1998)\citenamefont
  {Madison}, \citenamefont {Fischer}, \citenamefont {Diener}, \citenamefont
  {Niu},\ and\ \citenamefont {Raizen}}]{Madison1998}%
  \BibitemOpen
  \bibfield  {author} {\bibinfo {author} {\bibfnamefont {K.~W.}\ \bibnamefont
  {Madison}}, \bibinfo {author} {\bibfnamefont {M.~C.}\ \bibnamefont
  {Fischer}}, \bibinfo {author} {\bibfnamefont {R.~B.}\ \bibnamefont {Diener}},
  \bibinfo {author} {\bibfnamefont {Q.}~\bibnamefont {Niu}}, \ and\ \bibinfo
  {author} {\bibfnamefont {M.~G.}\ \bibnamefont {Raizen}},\ }\href@noop {}
  {\bibfield  {journal} {\bibinfo  {journal} {Physical Review Letters}\
  }\textbf {\bibinfo {volume} {81}},\ \bibinfo {pages} {5093} (\bibinfo {year}
  {1998})}\BibitemShut {NoStop}%
\bibitem [{\citenamefont {Szameit}\ \emph {et~al.}(2010)\citenamefont
  {Szameit}, \citenamefont {Garanovich}, \citenamefont {Heinrich},
  \citenamefont {Sukhorukov}, \citenamefont {Dreisow}, \citenamefont {Pertsch},
  \citenamefont {Nolte}, \citenamefont {Tünnermann}, \citenamefont {Longhi},\
  and\ \citenamefont {Kivshar}}]{Szameit2010}%
  \BibitemOpen
  \bibfield  {author} {\bibinfo {author} {\bibfnamefont {A.}~\bibnamefont
  {Szameit}}, \bibinfo {author} {\bibfnamefont {I.~L.}\ \bibnamefont
  {Garanovich}}, \bibinfo {author} {\bibfnamefont {M.}~\bibnamefont
  {Heinrich}}, \bibinfo {author} {\bibfnamefont {A.~A.}\ \bibnamefont
  {Sukhorukov}}, \bibinfo {author} {\bibfnamefont {F.}~\bibnamefont {Dreisow}},
  \bibinfo {author} {\bibfnamefont {T.}~\bibnamefont {Pertsch}}, \bibinfo
  {author} {\bibfnamefont {S.}~\bibnamefont {Nolte}}, \bibinfo {author}
  {\bibfnamefont {A.}~\bibnamefont {Tünnermann}}, \bibinfo {author}
  {\bibfnamefont {S.}~\bibnamefont {Longhi}}, \ and\ \bibinfo {author}
  {\bibfnamefont {Y.~S.}\ \bibnamefont {Kivshar}},\ }\href@noop {} {\bibfield
  {journal} {\bibinfo  {journal} {Physical Review Letters}\ }\textbf {\bibinfo
  {volume} {104}},\ \bibinfo {pages} {223903} (\bibinfo {year}
  {2010})}\BibitemShut {NoStop}%
\bibitem [{\citenamefont {Eckardt}\ \emph {et~al.}(2005)\citenamefont
  {Eckardt}, \citenamefont {Weiss},\ and\ \citenamefont
  {Holthaus}}]{Eckardt2005}%
  \BibitemOpen
  \bibfield  {author} {\bibinfo {author} {\bibfnamefont {A.}~\bibnamefont
  {Eckardt}}, \bibinfo {author} {\bibfnamefont {C.}~\bibnamefont {Weiss}}, \
  and\ \bibinfo {author} {\bibfnamefont {M.}~\bibnamefont {Holthaus}},\
  }\href@noop {} {\bibfield  {journal} {\bibinfo  {journal} {Physical Review
  Letters}\ }\textbf {\bibinfo {volume} {95}},\ \bibinfo {pages} {260404}
  (\bibinfo {year} {2005})}\BibitemShut {NoStop}%
\bibitem [{\citenamefont {Lignier}\ \emph {et~al.}(2007)\citenamefont
  {Lignier}, \citenamefont {Sias}, \citenamefont {Ciampini}, \citenamefont
  {Singh}, \citenamefont {Zenesini}, \citenamefont {Morsch},\ and\
  \citenamefont {Arimondo}}]{Lignier2007}%
  \BibitemOpen
  \bibfield  {author} {\bibinfo {author} {\bibfnamefont {H.}~\bibnamefont
  {Lignier}}, \bibinfo {author} {\bibfnamefont {C.}~\bibnamefont {Sias}},
  \bibinfo {author} {\bibfnamefont {D.}~\bibnamefont {Ciampini}}, \bibinfo
  {author} {\bibfnamefont {Y.}~\bibnamefont {Singh}}, \bibinfo {author}
  {\bibfnamefont {A.}~\bibnamefont {Zenesini}}, \bibinfo {author}
  {\bibfnamefont {O.}~\bibnamefont {Morsch}}, \ and\ \bibinfo {author}
  {\bibfnamefont {E.}~\bibnamefont {Arimondo}},\ }\href@noop {} {\bibfield
  {journal} {\bibinfo  {journal} {Physical Review Letters}\ }\textbf {\bibinfo
  {volume} {99}},\ \bibinfo {pages} {220403} (\bibinfo {year}
  {2007})}\BibitemShut {NoStop}%
\bibitem [{\citenamefont {Zenesini}\ \emph {et~al.}(2009)\citenamefont
  {Zenesini}, \citenamefont {Lignier}, \citenamefont {Ciampini}, \citenamefont
  {Morsch},\ and\ \citenamefont {Arimondo}}]{Zenesini2009}%
  \BibitemOpen
  \bibfield  {author} {\bibinfo {author} {\bibfnamefont {A.}~\bibnamefont
  {Zenesini}}, \bibinfo {author} {\bibfnamefont {H.}~\bibnamefont {Lignier}},
  \bibinfo {author} {\bibfnamefont {D.}~\bibnamefont {Ciampini}}, \bibinfo
  {author} {\bibfnamefont {O.}~\bibnamefont {Morsch}}, \ and\ \bibinfo {author}
  {\bibfnamefont {E.}~\bibnamefont {Arimondo}},\ }\href@noop {} {\bibfield
  {journal} {\bibinfo  {journal} {Physical Review Letters}\ }\textbf {\bibinfo
  {volume} {102}},\ \bibinfo {pages} {100403} (\bibinfo {year}
  {2009})}\BibitemShut {NoStop}%
\bibitem [{\citenamefont {Struck}\ \emph {et~al.}(2011)\citenamefont {Struck},
  \citenamefont {\"Oschl\"ager}, \citenamefont {Le~Targat}, \citenamefont
  {Soltan-Panahi}, \citenamefont {Eckardt}, \citenamefont {Lewenstein},
  \citenamefont {Windpassinger},\ and\ \citenamefont {Sengstock}}]{Struck2011}%
  \BibitemOpen
  \bibfield  {author} {\bibinfo {author} {\bibfnamefont {J.}~\bibnamefont
  {Struck}}, \bibinfo {author} {\bibfnamefont {C.}~\bibnamefont
  {\"Oschl\"ager}}, \bibinfo {author} {\bibfnamefont {R.}~\bibnamefont
  {Le~Targat}}, \bibinfo {author} {\bibfnamefont {P.}~\bibnamefont
  {Soltan-Panahi}}, \bibinfo {author} {\bibfnamefont {A.}~\bibnamefont
  {Eckardt}}, \bibinfo {author} {\bibfnamefont {M.}~\bibnamefont {Lewenstein}},
  \bibinfo {author} {\bibfnamefont {P.}~\bibnamefont {Windpassinger}}, \ and\
  \bibinfo {author} {\bibfnamefont {K.}~\bibnamefont {Sengstock}},\ }\href@noop
  {} {\bibfield  {journal} {\bibinfo  {journal} {Science}\ }\textbf {\bibinfo
  {volume} {333}},\ \bibinfo {pages} {996} (\bibinfo {year}
  {2011})}\BibitemShut {NoStop}%
\bibitem [{\citenamefont {Jiang}\ \emph {et~al.}(2011)\citenamefont {Jiang},
  \citenamefont {Kitagawa}, \citenamefont {Alicea}, \citenamefont {Akhmerov},
  \citenamefont {Pekker}, \citenamefont {Refael}, \citenamefont {Cirac},
  \citenamefont {Demler}, \citenamefont {Lukin},\ and\ \citenamefont
  {Zoller}}]{Jiang2011}%
  \BibitemOpen
  \bibfield  {author} {\bibinfo {author} {\bibfnamefont {L.}~\bibnamefont
  {Jiang}}, \bibinfo {author} {\bibfnamefont {T.}~\bibnamefont {Kitagawa}},
  \bibinfo {author} {\bibfnamefont {J.}~\bibnamefont {Alicea}}, \bibinfo
  {author} {\bibfnamefont {A.~R.}\ \bibnamefont {Akhmerov}}, \bibinfo {author}
  {\bibfnamefont {D.}~\bibnamefont {Pekker}}, \bibinfo {author} {\bibfnamefont
  {G.}~\bibnamefont {Refael}}, \bibinfo {author} {\bibfnamefont {J.~I.}\
  \bibnamefont {Cirac}}, \bibinfo {author} {\bibfnamefont {E.}~\bibnamefont
  {Demler}}, \bibinfo {author} {\bibfnamefont {M.~D.}\ \bibnamefont {Lukin}}, \
  and\ \bibinfo {author} {\bibfnamefont {P.}~\bibnamefont {Zoller}},\
  }\href@noop {} {\bibfield  {journal} {\bibinfo  {journal} {Physical Review
  Letters}\ }\textbf {\bibinfo {volume} {106}},\ \bibinfo {pages} {220402}
  (\bibinfo {year} {2011})}\BibitemShut {NoStop}%
\bibitem [{\citenamefont {Liu}\ \emph {et~al.}(2013)\citenamefont {Liu},
  \citenamefont {Levchenko},\ and\ \citenamefont {Baranger}}]{Liu2013}%
  \BibitemOpen
  \bibfield  {author} {\bibinfo {author} {\bibfnamefont {D.~E.}\ \bibnamefont
  {Liu}}, \bibinfo {author} {\bibfnamefont {A.}~\bibnamefont {Levchenko}}, \
  and\ \bibinfo {author} {\bibfnamefont {H.~U.}\ \bibnamefont {Baranger}},\
  }\href@noop {} {\bibfield  {journal} {\bibinfo  {journal} {Physical Review
  Letters}\ }\textbf {\bibinfo {volume} {111}},\ \bibinfo {pages} {047002}
  (\bibinfo {year} {2013})}\BibitemShut {NoStop}%
\bibitem [{\citenamefont {Kitagawa}\ \emph {et~al.}(2010)\citenamefont
  {Kitagawa}, \citenamefont {Berg}, \citenamefont {Rudner},\ and\ \citenamefont
  {Demler}}]{Kitagawa2010}%
  \BibitemOpen
  \bibfield  {author} {\bibinfo {author} {\bibfnamefont {T.}~\bibnamefont
  {Kitagawa}}, \bibinfo {author} {\bibfnamefont {E.}~\bibnamefont {Berg}},
  \bibinfo {author} {\bibfnamefont {M.}~\bibnamefont {Rudner}}, \ and\ \bibinfo
  {author} {\bibfnamefont {E.}~\bibnamefont {Demler}},\ }\href@noop {}
  {\bibfield  {journal} {\bibinfo  {journal} {Physical Review B}\ }\textbf
  {\bibinfo {volume} {82}},\ \bibinfo {pages} {235114} (\bibinfo {year}
  {2010})}\BibitemShut {NoStop}%
\bibitem [{\citenamefont {Lindner}\ \emph {et~al.}(2011)\citenamefont
  {Lindner}, \citenamefont {Refael},\ and\ \citenamefont
  {Galitski}}]{Lindner2011}%
  \BibitemOpen
  \bibfield  {author} {\bibinfo {author} {\bibfnamefont {N.~H.}\ \bibnamefont
  {Lindner}}, \bibinfo {author} {\bibfnamefont {G.}~\bibnamefont {Refael}}, \
  and\ \bibinfo {author} {\bibfnamefont {V.}~\bibnamefont {Galitski}},\
  }\href@noop {} {\bibfield  {journal} {\bibinfo  {journal} {Nat Phys}\
  }\textbf {\bibinfo {volume} {7}},\ \bibinfo {pages} {490} (\bibinfo {year}
  {2011})}\BibitemShut {NoStop}%
\bibitem [{\citenamefont {Hauke}\ \emph {et~al.}(2012)\citenamefont {Hauke},
  \citenamefont {Tieleman}, \citenamefont {Celi}, \citenamefont {\"Olschl\"ager},
  \citenamefont {Simonet}, \citenamefont {Struck}, \citenamefont {Weinberg},
  \citenamefont {Windpassinger}, \citenamefont {Sengstock}, \citenamefont
  {Lewenstein},\ and\ \citenamefont {Eckardt}}]{Hauke2012}%
  \BibitemOpen
  \bibfield  {author} {\bibinfo {author} {\bibfnamefont {P.}~\bibnamefont
  {Hauke}}, \bibinfo {author} {\bibfnamefont {O.}~\bibnamefont {Tieleman}},
  \bibinfo {author} {\bibfnamefont {A.}~\bibnamefont {Celi}}, \bibinfo {author}
  {\bibfnamefont {C.}~\bibnamefont {Ölschläger}}, \bibinfo {author}
  {\bibfnamefont {J.}~\bibnamefont {Simonet}}, \bibinfo {author} {\bibfnamefont
  {J.}~\bibnamefont {Struck}}, \bibinfo {author} {\bibfnamefont
  {M.}~\bibnamefont {Weinberg}}, \bibinfo {author} {\bibfnamefont
  {P.}~\bibnamefont {Windpassinger}}, \bibinfo {author} {\bibfnamefont
  {K.}~\bibnamefont {Sengstock}}, \bibinfo {author} {\bibfnamefont
  {M.}~\bibnamefont {Lewenstein}}, \ and\ \bibinfo {author} {\bibfnamefont
  {A.}~\bibnamefont {Eckardt}},\ }\href@noop {} {\bibfield  {journal} {\bibinfo
   {journal} {Physical Review Letters}\ }\textbf {\bibinfo {volume} {109}},\
  \bibinfo {pages} {145301} (\bibinfo {year} {2012})}\BibitemShut {NoStop}%
\bibitem [{\citenamefont {Struck}\ \emph {et~al.}(2012)\citenamefont {Struck},
  \citenamefont {\"Oschl\"ager}, \citenamefont {Weinberg}, \citenamefont
  {Hauke}, \citenamefont {Simonet}, \citenamefont {Eckardt}, \citenamefont
  {Lewenstein}, \citenamefont {Sengstock},\ and\ \citenamefont
  {Windpassinger}}]{Struck2012}%
  \BibitemOpen
  \bibfield  {author} {\bibinfo {author} {\bibfnamefont {J.}~\bibnamefont
  {Struck}}, \bibinfo {author} {\bibfnamefont {C.}~\bibnamefont
  {\"Oschl\"ager}}, \bibinfo {author} {\bibfnamefont {M.}~\bibnamefont
  {Weinberg}}, \bibinfo {author} {\bibfnamefont {P.}~\bibnamefont {Hauke}},
  \bibinfo {author} {\bibfnamefont {J.}~\bibnamefont {Simonet}}, \bibinfo
  {author} {\bibfnamefont {A.}~\bibnamefont {Eckardt}}, \bibinfo {author}
  {\bibfnamefont {M.}~\bibnamefont {Lewenstein}}, \bibinfo {author}
  {\bibfnamefont {K.}~\bibnamefont {Sengstock}}, \ and\ \bibinfo {author}
  {\bibfnamefont {P.}~\bibnamefont {Windpassinger}},\ }\href@noop {} {\bibfield
   {journal} {\bibinfo  {journal} {Physical Review Letters}\ }\textbf {\bibinfo
  {volume} {108}},\ \bibinfo {pages} {225304} (\bibinfo {year}
  {2012})}\BibitemShut {NoStop}%
\bibitem [{\citenamefont {Aidelsburger}\ \emph {et~al.}(2013)\citenamefont
  {Aidelsburger}, \citenamefont {Atala}, \citenamefont {Lohse}, \citenamefont
  {Barreiro}, \citenamefont {Paredes},\ and\ \citenamefont
  {Bloch}}]{Aidelsburger2013}%
  \BibitemOpen
  \bibfield  {author} {\bibinfo {author} {\bibfnamefont {M.}~\bibnamefont
  {Aidelsburger}}, \bibinfo {author} {\bibfnamefont {M.}~\bibnamefont {Atala}},
  \bibinfo {author} {\bibfnamefont {M.}~\bibnamefont {Lohse}}, \bibinfo
  {author} {\bibfnamefont {J.~T.}\ \bibnamefont {Barreiro}}, \bibinfo {author}
  {\bibfnamefont {B.}~\bibnamefont {Paredes}}, \ and\ \bibinfo {author}
  {\bibfnamefont {I.}~\bibnamefont {Bloch}},\ }\href@noop {} {\bibfield
  {journal} {\bibinfo  {journal} {Physical Review Letters}\ }\textbf {\bibinfo
  {volume} {111}},\ \bibinfo {pages} {185301} (\bibinfo {year}
  {2013})}\BibitemShut {NoStop}%
\bibitem [{\citenamefont {Miyake}\ \emph {et~al.}(2013)\citenamefont {Miyake},
  \citenamefont {Siviloglou}, \citenamefont {Kennedy}, \citenamefont {Burton},\
  and\ \citenamefont {Ketterle}}]{Miyake2013}%
  \BibitemOpen
  \bibfield  {author} {\bibinfo {author} {\bibfnamefont {H.}~\bibnamefont
  {Miyake}}, \bibinfo {author} {\bibfnamefont {G.~A.}\ \bibnamefont
  {Siviloglou}}, \bibinfo {author} {\bibfnamefont {C.~J.}\ \bibnamefont
  {Kennedy}}, \bibinfo {author} {\bibfnamefont {W.~C.}\ \bibnamefont {Burton}},
  \ and\ \bibinfo {author} {\bibfnamefont {W.}~\bibnamefont {Ketterle}},\
  }\href@noop {} {\bibfield  {journal} {\bibinfo  {journal} {Physical Review
  Letters}\ }\textbf {\bibinfo {volume} {111}},\ \bibinfo {pages} {185302}
  (\bibinfo {year} {2013})}\BibitemShut {NoStop}%
\bibitem [{\citenamefont {Goldman}\ and\ \citenamefont
  {Dalibard}(2014)}]{Goldman2014}%
  \BibitemOpen
  \bibfield  {author} {\bibinfo {author} {\bibfnamefont {N.}~\bibnamefont
  {Goldman}}\ and\ \bibinfo {author} {\bibfnamefont {J.}~\bibnamefont
  {Dalibard}},\ }\href@noop {} {\bibfield  {journal} {\bibinfo  {journal}
  {Physical Review X}\ }\textbf {\bibinfo {volume} {4}},\ \bibinfo {pages}
  {031027} (\bibinfo {year} {2014})}\BibitemShut {NoStop}%
\bibitem [{\citenamefont {G\'omez-Le\'on}\ and\ \citenamefont
  {Platero}(2013)}]{Gomez-Leon2013}%
  \BibitemOpen
  \bibfield  {author} {\bibinfo {author} {\bibfnamefont {A.}~\bibnamefont
  {G\'omez-Le\'on}}\ and\ \bibinfo {author} {\bibfnamefont {G.}~\bibnamefont
  {Platero}},\ }\href@noop {} {\bibfield  {journal} {\bibinfo  {journal}
  {Physical Review Letters}\ }\textbf {\bibinfo {volume} {110}},\ \bibinfo
  {pages} {200403} (\bibinfo {year} {2013})}\BibitemShut {NoStop}%
\bibitem [{\citenamefont {Rudner}\ \emph {et~al.}(2013)\citenamefont {Rudner},
  \citenamefont {Lindner}, \citenamefont {Berg},\ and\ \citenamefont
  {Levin}}]{Rudner2013}%
  \BibitemOpen
  \bibfield  {author} {\bibinfo {author} {\bibfnamefont {M.~S.}\ \bibnamefont
  {Rudner}}, \bibinfo {author} {\bibfnamefont {N.~H.}\ \bibnamefont {Lindner}},
  \bibinfo {author} {\bibfnamefont {E.}~\bibnamefont {Berg}}, \ and\ \bibinfo
  {author} {\bibfnamefont {M.}~\bibnamefont {Levin}},\ }\href@noop {}
  {\bibfield  {journal} {\bibinfo  {journal} {Physical Review X}\ }\textbf
  {\bibinfo {volume} {3}},\ \bibinfo {pages} {031005} (\bibinfo {year}
  {2013})}\BibitemShut {NoStop}%
\bibitem [{\citenamefont {Novoselov}\ \emph {et~al.}(2005)\citenamefont
  {Novoselov}, \citenamefont {Geim}, \citenamefont {Morozov}, \citenamefont
  {Jiang}, \citenamefont {Katsnelson}, \citenamefont {Grigorieva},
  \citenamefont {Dubonos},\ and\ \citenamefont {Firsov}}]{Novoselov2005}%
  \BibitemOpen
  \bibfield  {author} {\bibinfo {author} {\bibfnamefont {K.~S.}\ \bibnamefont
  {Novoselov}}, \bibinfo {author} {\bibfnamefont {A.~K.}\ \bibnamefont {Geim}},
  \bibinfo {author} {\bibfnamefont {S.~V.}\ \bibnamefont {Morozov}}, \bibinfo
  {author} {\bibfnamefont {D.}~\bibnamefont {Jiang}}, \bibinfo {author}
  {\bibfnamefont {M.~I.}\ \bibnamefont {Katsnelson}}, \bibinfo {author}
  {\bibfnamefont {I.~V.}\ \bibnamefont {Grigorieva}}, \bibinfo {author}
  {\bibfnamefont {S.~V.}\ \bibnamefont {Dubonos}}, \ and\ \bibinfo {author}
  {\bibfnamefont {A.~A.}\ \bibnamefont {Firsov}},\ }\href@noop {} {\bibfield
  {journal} {\bibinfo  {journal} {Nature}\ }\textbf {\bibinfo {volume} {438}},\
  \bibinfo {pages} {197} (\bibinfo {year} {2005})}\BibitemShut {NoStop}%
\bibitem [{\citenamefont {Castro~Neto}\ \emph {et~al.}(2009)\citenamefont
  {Castro~Neto}, \citenamefont {Guinea}, \citenamefont {Peres}, \citenamefont
  {Novoselov},\ and\ \citenamefont {Geim}}]{CastroNeto2009}%
  \BibitemOpen
  \bibfield  {author} {\bibinfo {author} {\bibfnamefont {A.~H.}\ \bibnamefont
  {Castro~Neto}}, \bibinfo {author} {\bibfnamefont {F.}~\bibnamefont {Guinea}},
  \bibinfo {author} {\bibfnamefont {N.~M.~R.}\ \bibnamefont {Peres}}, \bibinfo
  {author} {\bibfnamefont {K.~S.}\ \bibnamefont {Novoselov}}, \ and\ \bibinfo
  {author} {\bibfnamefont {A.~K.}\ \bibnamefont {Geim}},\ }\href@noop {}
  {\bibfield  {journal} {\bibinfo  {journal} {Reviews of Modern Physics}\
  }\textbf {\bibinfo {volume} {81}},\ \bibinfo {pages} {109} (\bibinfo {year}
  {2009})}\BibitemShut {NoStop}%
\bibitem [{\citenamefont {Shapere}\ and\ \citenamefont
  {Wilczek}(2012)}]{Shapere2012}%
  \BibitemOpen
  \bibfield  {author} {\bibinfo {author} {\bibfnamefont {A.}~\bibnamefont
  {Shapere}}\ and\ \bibinfo {author} {\bibfnamefont {F.}~\bibnamefont
  {Wilczek}},\ }\href@noop {} {\bibfield  {journal} {\bibinfo  {journal}
  {Physical Review Letters}\ }\textbf {\bibinfo {volume} {109}},\ \bibinfo
  {pages} {160402} (\bibinfo {year} {2012})}\BibitemShut {NoStop}%
\bibitem [{\citenamefont {Wilczek}(2012)}]{Wilczek2012}%
  \BibitemOpen
  \bibfield  {author} {\bibinfo {author} {\bibfnamefont {F.}~\bibnamefont
  {Wilczek}},\ }\href@noop {} {\bibfield  {journal} {\bibinfo  {journal}
  {Physical Review Letters}\ }\textbf {\bibinfo {volume} {109}},\ \bibinfo
  {pages} {160401} (\bibinfo {year} {2012})}\BibitemShut {NoStop}%
\bibitem [{\citenamefont {Li}\ \emph {et~al.}(2012)\citenamefont {Li},
  \citenamefont {Gong}, \citenamefont {Yin}, \citenamefont {Quan},
  \citenamefont {Yin}, \citenamefont {Zhang}, \citenamefont {Duan},\ and\
  \citenamefont {Zhang}}]{Li2012a}%
  \BibitemOpen
  \bibfield  {author} {\bibinfo {author} {\bibfnamefont {T.}~\bibnamefont
  {Li}}, \bibinfo {author} {\bibfnamefont {Z.-X.}\ \bibnamefont {Gong}},
  \bibinfo {author} {\bibfnamefont {Z.-Q.}\ \bibnamefont {Yin}}, \bibinfo
  {author} {\bibfnamefont {H.~T.}\ \bibnamefont {Quan}}, \bibinfo {author}
  {\bibfnamefont {X.}~\bibnamefont {Yin}}, \bibinfo {author} {\bibfnamefont
  {P.}~\bibnamefont {Zhang}}, \bibinfo {author} {\bibfnamefont {L.~M.}\
  \bibnamefont {Duan}}, \ and\ \bibinfo {author} {\bibfnamefont
  {X.}~\bibnamefont {Zhang}},\ }\href@noop {} {\bibfield  {journal} {\bibinfo
  {journal} {Physical Review Letters}\ }\textbf {\bibinfo {volume} {109}},\
  \bibinfo {pages} {163001} (\bibinfo {year} {2012})}\BibitemShut {NoStop}%
\bibitem [{\citenamefont {Guo}\ \emph {et~al.}(2013)\citenamefont {Guo},
  \citenamefont {Marthaler},\ and\ \citenamefont {Sch\"on}}]{Guo2013}%
  \BibitemOpen
  \bibfield  {author} {\bibinfo {author} {\bibfnamefont {L.}~\bibnamefont
  {Guo}}, \bibinfo {author} {\bibfnamefont {M.}~\bibnamefont {Marthaler}}, \
  and\ \bibinfo {author} {\bibfnamefont {G.}~\bibnamefont {Sch\"on}},\
  }\href@noop {} {\bibfield  {journal} {\bibinfo  {journal} {Physical Review
  Letters}\ }\textbf {\bibinfo {volume} {111}},\ \bibinfo {pages} {205303}
  (\bibinfo {year} {2013})}\BibitemShut {NoStop}%
\bibitem [{\citenamefont {Cooper}\ and\ \citenamefont
  {Dalibard}(2013)}]{Cooper2013}%
  \BibitemOpen
  \bibfield  {author} {\bibinfo {author} {\bibfnamefont {N.~R.}\ \bibnamefont
  {Cooper}}\ and\ \bibinfo {author} {\bibfnamefont {J.}~\bibnamefont
  {Dalibard}},\ }\href@noop {} {\bibfield  {journal} {\bibinfo  {journal}
  {Physical Review Letters}\ }\textbf {\bibinfo {volume} {110}},\ \bibinfo
  {pages} {185301} (\bibinfo {year} {2013})}\BibitemShut {NoStop}%
\bibitem [{\citenamefont {Lamb}(1939)}]{Lamb1939}%
  \BibitemOpen
  \bibfield  {author} {\bibinfo {author} {\bibfnamefont {W.~E.}\ \bibnamefont
  {Lamb}},\ }\href@noop {} {\bibfield  {journal} {\bibinfo  {journal} {Physical
  Review}\ }\textbf {\bibinfo {volume} {55}},\ \bibinfo {pages} {190} (\bibinfo
  {year} {1939})}\BibitemShut {NoStop}%
\bibitem [{\citenamefont {M\"ossbauer}(1958)}]{Mossbauer1958}%
  \BibitemOpen
  \bibfield  {author} {\bibinfo {author} {\bibfnamefont {R.}~\bibnamefont
  {M\"ossbauer}},\ }\href@noop {} {\bibfield  {journal} {\bibinfo  {journal}
  {Zeitschrift f\"ur Physik}\ }\textbf {\bibinfo {volume} {151}},\ \bibinfo
  {pages} {124} (\bibinfo {year} {1958})}\BibitemShut {NoStop}%
\bibitem [{\citenamefont {Frauenfelder}(1962)}]{Frauenfelder1962}%
  \BibitemOpen
  \bibfield  {author} {\bibinfo {author} {\bibfnamefont {H.}~\bibnamefont
  {Frauenfelder}},\ }\href@noop {} {\emph {\bibinfo {title} {The Mossbauer
  effect a review, with a collection of reprints}}}\ (\bibinfo  {publisher}
  {New York, W.A. Benjamin},\ \bibinfo {address} {New York},\ \bibinfo {year}
  {1962})\BibitemShut {NoStop}%
\bibitem [{\citenamefont {Dicke}(1954)}]{Dicke1954}%
  \BibitemOpen
  \bibfield  {author} {\bibinfo {author} {\bibfnamefont {R.~H.}\ \bibnamefont
  {Dicke}},\ }\href@noop {} {\bibfield  {journal} {\bibinfo  {journal}
  {Physical Review}\ }\textbf {\bibinfo {volume} {93}},\ \bibinfo {pages} {99}
  (\bibinfo {year} {1954})}\BibitemShut {NoStop}%
\bibitem [{\citenamefont {Scully}\ \emph {et~al.}(2006)\citenamefont {Scully},
  \citenamefont {Fry}, \citenamefont {Ooi},\ and\ \citenamefont
  {Wódkiewicz}}]{Scully2006}%
  \BibitemOpen
  \bibfield  {author} {\bibinfo {author} {\bibfnamefont {M.~O.}\ \bibnamefont
  {Scully}}, \bibinfo {author} {\bibfnamefont {E.~S.}\ \bibnamefont {Fry}},
  \bibinfo {author} {\bibfnamefont {C.~H.~R.}\ \bibnamefont {Ooi}}, \ and\
  \bibinfo {author} {\bibfnamefont {K.}~\bibnamefont {Wódkiewicz}},\
  }\href@noop {} {\bibfield  {journal} {\bibinfo  {journal} {Physical Review
  Letters}\ }\textbf {\bibinfo {volume} {96}},\ \bibinfo {pages} {010501}
  (\bibinfo {year} {2006})}\BibitemShut {NoStop}%
\bibitem [{\citenamefont {Artoni}\ and\ \citenamefont
  {La~Rocca}(2006)}]{Artoni2006}%
  \BibitemOpen
  \bibfield  {author} {\bibinfo {author} {\bibfnamefont {M.}~\bibnamefont
  {Artoni}}\ and\ \bibinfo {author} {\bibfnamefont {G.~C.}\ \bibnamefont
  {La~Rocca}},\ }\href@noop {} {\bibfield  {journal} {\bibinfo  {journal}
  {Physical Review Letters}\ }\textbf {\bibinfo {volume} {96}},\ \bibinfo
  {pages} {073905} (\bibinfo {year} {2006})}\BibitemShut {NoStop}%
\bibitem [{Sup()}]{Supplementary}%
  \BibitemOpen
  \href@noop {} {\bibinfo  {journal} {See Supplemental Material at http://link.aps.org/
  supplemental/10.1103/PhysRevLett.114.043602, which
  includes Refs. [44-49]}\ }\BibitemShut {NoStop}%
\bibitem [{\citenamefont {Kapitza}\ and\ \citenamefont
  {Dirac}(1933)}]{Kapitza1933}%
  \BibitemOpen
\bibfield  {journal} {  }\bibfield  {author} {\bibinfo {author} {\bibfnamefont
  {P.~L.}\ \bibnamefont {Kapitza}}\ and\ \bibinfo {author} {\bibnamefont
  {Dirac}},\ }\href@noop {} {\bibfield  {journal} {\bibinfo  {journal}
  {Mathematical Proceedings of the Cambridge Philosophical Society}\ }\textbf
  {\bibinfo {volume} {29}},\ \bibinfo {pages} {297} (\bibinfo {year}
  {1933})}\BibitemShut {NoStop}%
\bibitem [{\citenamefont {Scully}(2009)}]{Scully2009}%
  \BibitemOpen
  \bibfield  {author} {\bibinfo {author} {\bibfnamefont {M.~O.}\ \bibnamefont
  {Scully}},\ }\href@noop {} {\bibfield  {journal} {\bibinfo  {journal}
  {Physical Review Letters}\ }\textbf {\bibinfo {volume} {102}},\ \bibinfo
  {pages} {143601} (\bibinfo {year} {2009})}\BibitemShut {NoStop}%
\bibitem [{\citenamefont {R\"ohlsberger}\ \emph {et~al.}(2010)\citenamefont
  {R\"ohlsberger}, \citenamefont {Schlage}, \citenamefont {Sahoo},
  \citenamefont {Couet},\ and\ \citenamefont {R\"uffer}}]{Rohlsberger2010}%
  \BibitemOpen
  \bibfield  {author} {\bibinfo {author} {\bibfnamefont {R.}~\bibnamefont
  {R\"ohlsberger}}, \bibinfo {author} {\bibfnamefont {K.}~\bibnamefont
  {Schlage}}, \bibinfo {author} {\bibfnamefont {B.}~\bibnamefont {Sahoo}},
  \bibinfo {author} {\bibfnamefont {S.}~\bibnamefont {Couet}}, \ and\ \bibinfo
  {author} {\bibfnamefont {R.}~\bibnamefont {R\"uffer}},\ }\href@noop {}
  {\bibfield  {journal} {\bibinfo  {journal} {Science}\ }\textbf {\bibinfo
  {volume} {328}},\ \bibinfo {pages} {1248} (\bibinfo {year}
  {2010})}\BibitemShut {NoStop}%
\bibitem [{\citenamefont {Feng}\ \emph {et~al.}(2014)\citenamefont {Feng},
  \citenamefont {Li},\ and\ \citenamefont {Zhu}}]{Feng2014}%
  \BibitemOpen
  \bibfield  {author} {\bibinfo {author} {\bibfnamefont {W.}~\bibnamefont
  {Feng}}, \bibinfo {author} {\bibfnamefont {Y.}~\bibnamefont {Li}}, \ and\
  \bibinfo {author} {\bibfnamefont {S.-Y.}\ \bibnamefont {Zhu}},\ }\href@noop
  {} {\bibfield  {journal} {\bibinfo  {journal} {Physical Review A}\ }\textbf
  {\bibinfo {volume} {89}},\ \bibinfo {pages} {013816} (\bibinfo {year}
  {2014})}\BibitemShut {NoStop}%
\bibitem [{\citenamefont {Anderson}(1958)}]{Anderson1958}%
  \BibitemOpen
  \bibfield  {author} {\bibinfo {author} {\bibfnamefont {P.~W.}\ \bibnamefont
  {Anderson}},\ }\href@noop {} {\bibfield  {journal} {\bibinfo  {journal}
  {Physical Review}\ }\textbf {\bibinfo {volume} {109}},\ \bibinfo {pages}
  {1492} (\bibinfo {year} {1958})}\BibitemShut {NoStop}%
\bibitem [{\citenamefont {Rivera}\ and\ \citenamefont
  {Schulz}(1999)}]{Rivera1999}%
  \BibitemOpen
  \bibfield  {author} {\bibinfo {author} {\bibfnamefont {P.~H.}\ \bibnamefont
  {Rivera}}\ and\ \bibinfo {author} {\bibfnamefont {P.~A.}\ \bibnamefont
  {Schulz}},\ }\href@noop {} {\bibfield  {journal} {\bibinfo  {journal}
  {Brazilian Journal of Physics}\ }\textbf {\bibinfo {volume} {29}},\ \bibinfo
  {pages} {685} (\bibinfo {year} {1999})}\BibitemShut {NoStop}%
\bibitem [{\citenamefont {Fano}(1961)}]{Fano1961}%
  \BibitemOpen
  \bibfield  {author} {\bibinfo {author} {\bibfnamefont {U.}~\bibnamefont
  {Fano}},\ }\href@noop {} {\bibfield  {journal} {\bibinfo  {journal} {Physical
  Review}\ }\textbf {\bibinfo {volume} {124}},\ \bibinfo {pages} {1866}
  (\bibinfo {year} {1961})}\BibitemShut {NoStop}%
\bibitem [{\citenamefont {Miroshnichenko}\ \emph {et~al.}(2010)\citenamefont
  {Miroshnichenko}, \citenamefont {Flach},\ and\ \citenamefont
  {Kivshar}}]{Miroshnichenko2010}%
  \BibitemOpen
  \bibfield  {author} {\bibinfo {author} {\bibfnamefont {A.~E.}\ \bibnamefont
  {Miroshnichenko}}, \bibinfo {author} {\bibfnamefont {S.}~\bibnamefont
  {Flach}}, \ and\ \bibinfo {author} {\bibfnamefont {Y.~S.}\ \bibnamefont
  {Kivshar}},\ }\href@noop {} {\bibfield  {journal} {\bibinfo  {journal}
  {Reviews of Modern Physics}\ }\textbf {\bibinfo {volume} {82}},\ \bibinfo
  {pages} {2257} (\bibinfo {year} {2010})}\BibitemShut {NoStop}%
\bibitem [{\citenamefont {Zak}(1993)}]{Zak1993}%
  \BibitemOpen
  \bibfield  {author} {\bibinfo {author} {\bibfnamefont {J.}~\bibnamefont
  {Zak}},\ }\href@noop {} {\bibfield  {journal} {\bibinfo  {journal} {Physical
  Review Letters}\ }\textbf {\bibinfo {volume} {71}},\ \bibinfo {pages} {2623}
  (\bibinfo {year} {1993})}\BibitemShut {NoStop}%
\bibitem [{\citenamefont {Zhao}(1991)}]{Zhao1991}%
  \BibitemOpen
  \bibfield  {author} {\bibinfo {author} {\bibfnamefont {X.-G.}\ \bibnamefont
  {Zhao}},\ }\href@noop {} {\bibfield  {journal} {\bibinfo  {journal} {Physics
  Letters A}\ }\textbf {\bibinfo {volume} {155}},\ \bibinfo {pages} {299}
  (\bibinfo {year} {1991})}\BibitemShut {NoStop}%
\bibitem [{\citenamefont {Holthaus}\ \emph {et~al.}(1995)\citenamefont
  {Holthaus}, \citenamefont {Ristow},\ and\ \citenamefont
  {Hone}}]{Holthaus1995}%
  \BibitemOpen
  \bibfield  {author} {\bibinfo {author} {\bibfnamefont {M.}~\bibnamefont
  {Holthaus}}, \bibinfo {author} {\bibfnamefont {G.~H.}\ \bibnamefont
  {Ristow}}, \ and\ \bibinfo {author} {\bibfnamefont {D.~W.}\ \bibnamefont
  {Hone}},\ }\href@noop {} {\bibfield  {journal} {\bibinfo  {journal} {Physical
  Review Letters}\ }\textbf {\bibinfo {volume} {75}},\ \bibinfo {pages} {3914}
  (\bibinfo {year} {1995})}\BibitemShut {NoStop}%
\bibitem [{\citenamefont {Liu}\ and\ \citenamefont {Zhu}(1999)}]{Liu1999}%
  \BibitemOpen
  \bibfield  {author} {\bibinfo {author} {\bibfnamefont {R.-B.}\ \bibnamefont
  {Liu}}\ and\ \bibinfo {author} {\bibfnamefont {B.-F.}\ \bibnamefont {Zhu}},\
  }\href@noop {} {\bibfield  {journal} {\bibinfo  {journal} {Physical Review
  B}\ }\textbf {\bibinfo {volume} {59}},\ \bibinfo {pages} {5759} (\bibinfo
  {year} {1999})}\BibitemShut {NoStop}%
\bibitem [{\citenamefont {Haldane}(1988)}]{Haldane1988}%
  \BibitemOpen
  \bibfield  {author} {\bibinfo {author} {\bibfnamefont {F.~D.~M.}\
  \bibnamefont {Haldane}},\ }\href@noop {} {\bibfield  {journal} {\bibinfo
  {journal} {Physical Review Letters}\ }\textbf {\bibinfo {volume} {61}},\
  \bibinfo {pages} {2015} (\bibinfo {year} {1988})}\BibitemShut {NoStop}%
\bibitem [{\citenamefont {Wang}\ \emph {et~al.}()\citenamefont {Wang},
  \citenamefont {Cai}, \citenamefont {Yuan}, \citenamefont {Liu},\ and\
  \citenamefont {Scully}}]{Wang2015}%
  \BibitemOpen
  \bibfield  {author} {\bibinfo {author} {\bibfnamefont {D.-W.}\ \bibnamefont
  {Wang}}, \bibinfo {author} {\bibfnamefont {H.}~\bibnamefont {Cai}}, \bibinfo
  {author} {\bibfnamefont {L.}~\bibnamefont {Yuan}}, \bibinfo {author}
  {\bibfnamefont {R.-B.}\ \bibnamefont {Liu}}, \ and\ \bibinfo {author}
  {\bibfnamefont {S.-Y.}~\bibnamefont {Zhu}},\
  }\href@noop {} {\bibinfo  {journal} {arXiv:1501.04099 [physics.optics]}\ }\BibitemShut {NoStop}%
\bibitem [{\citenamefont {Zhang}\ and\ \citenamefont {Hu}(2001)}]{Zhang2001}%
  \BibitemOpen
\bibfield  {journal} {  }\bibfield  {author} {\bibinfo {author} {\bibfnamefont
  {S.-C.}\ \bibnamefont {Zhang}}\ and\ \bibinfo {author} {\bibfnamefont
  {J.}~\bibnamefont {Hu}},\ }\href@noop {} {\bibfield  {journal} {\bibinfo
  {journal} {Science}\ }\textbf {\bibinfo {volume} {294}},\ \bibinfo {pages}
  {823} (\bibinfo {year} {2001})}\BibitemShut {NoStop}%
\bibitem [{\citenamefont {Ling}\ \emph {et~al.}(1998)\citenamefont {Ling},
  \citenamefont {Li},\ and\ \citenamefont {Xiao}}]{Ling1998}%
  \BibitemOpen
  \bibfield  {author} {\bibinfo {author} {\bibfnamefont {H.~Y.}\ \bibnamefont
  {Ling}}, \bibinfo {author} {\bibfnamefont {Y.-Q.}\ \bibnamefont {Li}}, \ and\
  \bibinfo {author} {\bibfnamefont {M.}~\bibnamefont {Xiao}},\ }\href@noop {}
  {\bibfield  {journal} {\bibinfo  {journal} {Physical Review A}\ }\textbf
  {\bibinfo {volume} {57}},\ \bibinfo {pages} {1338} (\bibinfo {year}
  {1998})}\BibitemShut {NoStop}%
\bibitem [{\citenamefont {Mitsunaga}\ and\ \citenamefont
  {Imoto}(1999)}]{Mitsunaga1999}%
  \BibitemOpen
  \bibfield  {author} {\bibinfo {author} {\bibfnamefont {M.}~\bibnamefont
  {Mitsunaga}}\ and\ \bibinfo {author} {\bibfnamefont {N.}~\bibnamefont
  {Imoto}},\ }\href@noop {} {\bibfield  {journal} {\bibinfo  {journal}
  {Physical Review A}\ }\textbf {\bibinfo {volume} {59}},\ \bibinfo {pages}
  {4773} (\bibinfo {year} {1999})}\BibitemShut {NoStop}%
\bibitem [{\citenamefont {Cardoso}\ and\ \citenamefont
  {Tabosa}(2002)}]{Cardoso2002}%
  \BibitemOpen
  \bibfield  {author} {\bibinfo {author} {\bibfnamefont {G.~C.}\ \bibnamefont
  {Cardoso}}\ and\ \bibinfo {author} {\bibfnamefont {J.~W.~R.}\ \bibnamefont
  {Tabosa}},\ }\href@noop {} {\bibfield  {journal} {\bibinfo  {journal}
  {Physical Review A}\ }\textbf {\bibinfo {volume} {65}},\ \bibinfo {pages}
  {033803} (\bibinfo {year} {2002})}\BibitemShut {NoStop}%
\bibitem [{\citenamefont {Wang}\ \emph {et~al.}(2003)\citenamefont {Wang},
  \citenamefont {Zhu}, \citenamefont {Jiang},\ and\ \citenamefont
  {Zhan}}]{Wang2003}%
  \BibitemOpen
  \bibfield  {author} {\bibinfo {author} {\bibfnamefont {J.}~\bibnamefont
  {Wang}}, \bibinfo {author} {\bibfnamefont {Y.}~\bibnamefont {Zhu}}, \bibinfo
  {author} {\bibfnamefont {K.~J.}\ \bibnamefont {Jiang}}, \ and\ \bibinfo
  {author} {\bibfnamefont {M.~S.}\ \bibnamefont {Zhan}},\ }\href@noop {}
  {\bibfield  {journal} {\bibinfo  {journal} {Physical Review A}\ }\textbf
  {\bibinfo {volume} {68}},\ \bibinfo {pages} {063810} (\bibinfo {year}
  {2003})}\BibitemShut {NoStop}%
\bibitem [{\citenamefont {{Wang}}\ \emph {et~al.}(2013)\citenamefont {{Wang}},
  \citenamefont {{Zhu}}, \citenamefont {{Evers}},\ and\ \citenamefont
  {{Scully}}}]{2013arXiv1305.3636W}%
  \BibitemOpen
  \bibfield  {author} {\bibinfo {author} {\bibfnamefont {D.-W.}\ \bibnamefont
  {{Wang}}}, \bibinfo {author} {\bibfnamefont {S.-Y.}\ \bibnamefont {{Zhu}}},
  \bibinfo {author} {\bibfnamefont {J.}~\bibnamefont {{Evers}}}, \ and\
  \bibinfo {author} {\bibfnamefont {M.~O.}\ \bibnamefont {{Scully}}},\
  }\href@noop {} 
  {\bibfield  {journal} {\bibinfo  {journal} {Physical Review A}\ }\textbf
  {\bibinfo {volume} {91}},\ \bibinfo {pages} {011801(R)} (\bibinfo {year}
  {2015})}\BibitemShut {NoStop}%
\bibitem [{\citenamefont {Wang}\ and\ \citenamefont {Scully}(2014)}]{Wang2014}%
  \BibitemOpen
  \bibfield  {author} {\bibinfo {author} {\bibfnamefont {D.-W.}\ \bibnamefont
  {Wang}}\ and\ \bibinfo {author} {\bibfnamefont {M.~O.}\ \bibnamefont
  {Scully}},\ }\href@noop {} {\bibfield  {journal} {\bibinfo  {journal}
  {Physical Review Letters}\ }\textbf {\bibinfo {volume} {113}},\ \bibinfo
  {pages} {083601} (\bibinfo {year} {2014})}\BibitemShut {NoStop}%
\bibitem [{\citenamefont {Wang}\ \emph {et~al.}(2013)\citenamefont {Wang},
  \citenamefont {Zhou}, \citenamefont {Guo}, \citenamefont {Zhang},
  \citenamefont {Evers},\ and\ \citenamefont {Zhu}}]{Wang2013}%
  \BibitemOpen
  \bibfield  {author} {\bibinfo {author} {\bibfnamefont {D.-W.}\ \bibnamefont
  {Wang}}, \bibinfo {author} {\bibfnamefont {H.-T.}\ \bibnamefont {Zhou}},
  \bibinfo {author} {\bibfnamefont {M.-J.}\ \bibnamefont {Guo}}, \bibinfo
  {author} {\bibfnamefont {J.-X.}\ \bibnamefont {Zhang}}, \bibinfo {author}
  {\bibfnamefont {J.}~\bibnamefont {Evers}}, \ and\ \bibinfo {author}
  {\bibfnamefont {S.-Y.}\ \bibnamefont {Zhu}},\ }\href@noop {} {\bibfield
  {journal} {\bibinfo  {journal} {Physical Review Letters}\ }\textbf {\bibinfo
  {volume} {110}},\ \bibinfo {pages} {093901} (\bibinfo {year}
  {2013})}\BibitemShut {NoStop}%
\end{thebibliography}

\end{document}